\documentclass[twocolumn,tighten]{aastex63}
\pdfoutput=1 
\usepackage{amsmath,amstext}
\usepackage{apjfonts}
\usepackage{afterpage}
\usepackage{xcolor}
\usepackage{threeparttable}
\usepackage{graphicx}
\usepackage{epsfig}
\usepackage{booktabs}
\usepackage{multirow}
\usepackage{float}
\usepackage{mathrsfs}
\DeclareUnicodeCharacter{1E11}{\k{a}}
\setwatermarkfontsize{100px}

\begin{document}
\author[0009-0003-6348-7143]{Tong Zhao}
\affiliation{School of Physics, Peking University, Beijing 100871, China}

\author[0009-0007-3817-8188]{Shunshun Cao}
\affiliation{School of Physics, Peking University, Beijing 100871, China}

\author[0000-0002-9042-3044]{Renxin Xu}
\affiliation{School of Physics, Peking University, Beijing 100871, China}

\title{Constraining Pulsar Radiative Geometry via Multi-wavelength Modeling}

\begin{abstract}
We propose that jointly modeling the thermal X-ray pulse profiles and the polarization position angles offers an effective means of locating the polarization orientation focus of the pulsar's coherent radiation. From the X-ray pulse-profile measurement we constrain the colatitude of the center of the thermal X-ray emission, which corresponds to the center of in-falling particles within the polar cap, while the RVM fitting yields the inclination angle of the focus point of polarization orientations. Thus, consistency between these two independent angle measurements would imply that the RVM fit faithfully recovers the inclination of the plasma flow center, and this center coincides with the polarization orientation focus. Conversely, the discrepancy would suggest that the polarization state of the radio emission changes as it propagates because the evolution of wave modes during wave propagation in the magnetosphere strongly depends on magnetic field orientations.
\end{abstract}

\section{Introduction}
Modeling the X-ray pulse profiles of rotation-powered millisecond pulsars that produce periodic thermal X-ray emission from hot polar caps to measure neutron star masses and radii is one of the key scientific objectives of the Neutron Star Interior Composition Explorer (NICER) and the enhanced X-ray Timing and Polarimetry (eXTP) missions~\citep{NICERins,eXTP2025}. In addition to masses and radii of neutron stars, other physics properties given by the parameter inference are also of special interest in some researches, such as the comparison between multi-messenger observations. For example, ~\citet{Petri2025} combined the latest NICER phase aligned thermal X-ray pulse-profile of PSR J0740+6620 with $\gamma$-ray pulse-profiles and radio polarization to deduce the best magnetic field configuration that can simultaneously reproduce the light curves in these respective bands.

In rotational powered pulsars, the observed thermal emission from their surfaces is thought to arise from bombardments of magnetospheric charges onto their polar caps~\citep{Ruderman1975, Arons1981}. The polar cap region is defined as the region on the neutron star surface where dipolar magnetic field lines are open, allowing charged particles to escape and generate the stars powerful electromagnetic emissions. Thus, the center of the polar cap is the the center of the open magnetic field lines. However, the thermal X-ray emission area is defined as the area bombarded by the in-falling particles, which is only a part of the polar cap region whose center is the center of the in-falling particle region~\citep{Hermsen2013,Szary2017,Sznajder2020}. Thus, the colatitude of the hot spot center measured by the thermal X-ray pulse profile modeling is the colatitude of the center of the in-falling particles rather than the center of the polar cap.

On the other hand, the rotating-vector model (RVM, \citealp{RVM1969}) is frequently used to fit the polarization angle of the radio emission. RVM fitting is believed to give us the angle between the magnetic moment and the rotation axis, i.e. the inclination angle, because it is assumed that the polarization of the coherent emission is parallel to the local magnetic field. But recent effort on simulation~\citep{Jan2026} has argued that the polarization of pulsar coherent radiation is centered at the center of plasma flow, which could be different from the center of dipole magnetic field. The wide usage of RVM fitting might be questionable.

However, \cite{Jan2026} does not consider the propagation effects in the pulsar magnetosphere. The evolution of wave modes during wave propagation in the magnetosphere strongly depends on magnetic field orientations~\citep[e.g.,][]{pl2000,petrova2001}. We speculate that if the propagation effects is strong enough, the resulting emission will still be modulated by the dipole magnetic field of the pulsar, even if it is purely ``non-RVM'' at the radiation altitude. Besides further discussions of emission physics in pulsar magnetospheres, we want some constraints from observations. For pulsars with pulsar wind nebulae (PWNe), multi-band observations have been made, arguing that for the RVM paradigm Vela pulsar, the outgoing waves are extraordinary modes~\citep{helfand2001}.

In this work, we propose that the comparison between the thermal X-ray pulse-profile measurement and the polarization position angle measurement is a promising way to address this problem. If the colatitude of the hot spot measured by the pulse-profile modeling is consistent with the magnetic inclination angle measured by RVM, the propagation effects do not matter, and the RVM is measuring the position of the center of the plasma flows in the polar cap rather than the position of the magnetic field center. If the two results are not the same, we can infer that the polarization state of the radio emission changes as it propagates.

Although the uncertainty of the hot spot colatitude measurement is usually too large for this purpose according to the previous results of NICER observations~\citep{Riley2019,Miller2019,Riley2021,Miller2021,Choudhury2024,Salmi2024,Mauviard2025}, future observations from the enhanced X-ray Timing and Polarimetry (eXTP) mission scheduled to launch in early 2030 may provide a good chance to measure this angle with high precision~\citep{eXTP2025}. Thus, we want to study how the uncertainty of the hot spot colatitude measurement varies with neutron star properties and whether this uncertainty can be reduced to a level that meets the requirement of this research objective. We generate synthetic data and estimate the uncertainty of X-ray and radio measurements in the most ideal situation.

In \S2 and in \S3, we generate synthetic data of polarization profile and thermal X-ray pulse-profiles to study the uncertainties of measurements based on RVM and the thermal X-ray pulse-profile modeling. In \S4, we summarize and discuss the possibility of determining the impact of propagation effects on the radio polarization position angle by comparing thermal X-ray pulse-profile measurements and polarization position angle measurements.

\section{Synthetic Data and Parameter Inference Results for Thermal X-ray Pulse-Profiles}\label{sec:xray}
We want to estimate whether the uncertainty of the spot collatitude measurement can be reduced to the level that can distinguish the current center and the magnetic axis. In the most ideal situation where a dipole magnetic field is assumed and the two hot spots are antipodal, there are at least six unknown parameters in the thermal X-ray pulse-profile model, the compactness $u=2GM/Rc^2$, the inclination angle of the observer $\zeta$, the colatitude of the spot $\alpha$, the effective temperature $T$, the overall factor proportional to the spot area over the distance square $A=dS/D^2$, and the hydrogen column density $N_H$ if the interstellar extinction is considered. Certainly, the background level is also important in the real observation, for the estimation of the uncertainty for the most ideal situation, we assume a low level background and ignore it. We simply generate synthetic pulse-profiles and add a Gaussian noise. The uncertainty of the spot colatitude measurement depends on all the six parameters. To minimize the uncertainty, we need to figure out proper values for them as the injected values for our synthetic pulse-profiles. The effective area of NICER peaks at about 1~keV. Thus, the temperature of the optimal blackbody spectrum that peaks at the same energy is about 0.5~keV. However, the effective temperature of the thermal X-ray spectrum for NICER sources are all about 0.1~keV according to the observational results~\citep{Riley2019,Miller2019,Riley2021,Miller2021,Choudhury2024,Salmi2024,Mauviard2025}. Thus, we assume that our effective temperature is 0.3~keV because it is probably difficult to find high temperature hot spots according to the previous observations. The spot area over the distance square gives an overall factor to the final flux. Thus, it influences the uncertainty by changing the total number of photons. We vary this factor to generate pulse-profiles with a total number of photons about $10^6$ for synthetic NICER data. For $u$, $\zeta$ and $\alpha$, we generated and fit a large number of synthetic pulse-profiles for different parameters and found a set of proper values for them: $u=0.32$, $\zeta=25^\circ$ and $\alpha=85^\circ$. This choice of parameters is not too extreme and can minimize the uncertainty of the spot colatitude measurement as much as possible. All parameters and their injected values for synthetic pulse-profiles are listed in Table~\ref{table:1}. Here, the overall normalization factor $A$ is scaled by a fiducial value $A_0=dS_0/D_0^2$ where $dS_0=10^6~\text{m}^2$ and $D_0=2$~kpc to avoid very small numbers.
\begin{table*}[t]
\caption{Free Parameters and their assumed values}
\begin{center}
\begin{tabular}{c c c c}
\toprule
Parameter & Description & Units & assumed value \\
\midrule
$u$ & Compactness & dimensionless & $0.32$~km \\ 
$\zeta$ & Observer inclination angle & degrees & $25$ \\
$\alpha$ & Colatitude of the primary spot& degrees & $85$ \\
$T$ & Effective temperature & keV & $0.15$\\
$A$ & $dS/D^2/A_0$ & dimensionless & $1$ \\ 
$N_H$ & Column Density of Hydrogen& $10^{20}$~cm$^{-2}$& 1.5\\
\bottomrule
\end{tabular}
\end{center}
\label{table:1}
\end{table*}

\begin{figure*}
	\centering
	\includegraphics[width=0.8\textwidth, keepaspectratio]{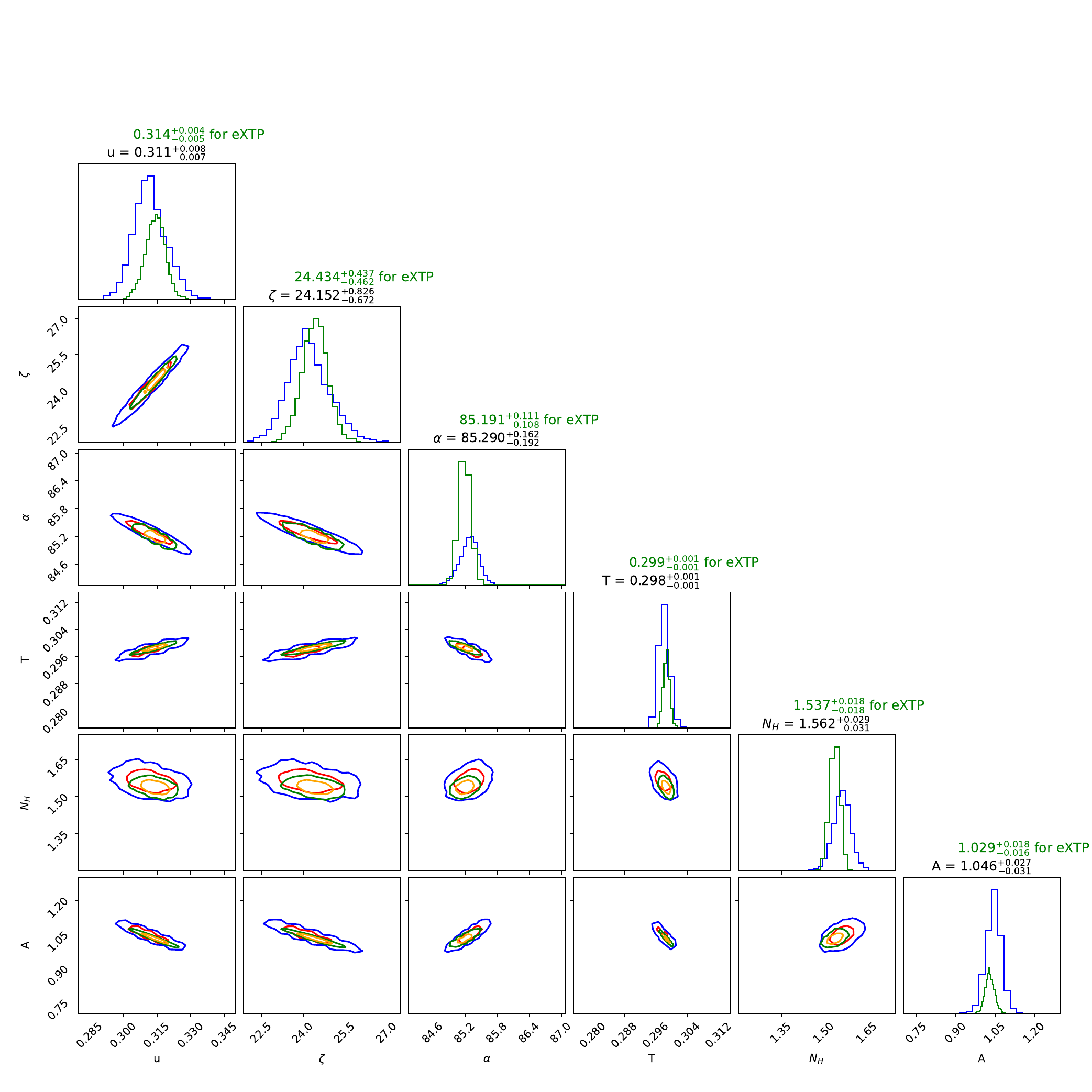}
	\caption{\label{fig:fig1} Fit results for synthetic NICER and eXTP data. 68-th and 95-th percentiles of the marginalized posterior distributions for synthetic NICER data are shown in red and blue, and for synthetic eXTP data in orange and green. The width of marginalized posterior over $\alpha$ for NICER and eXTP is about $0.2^\circ$ and $0.1^\circ$ respectively.}
\end{figure*}

For our synthetic data, we choose the photon energy range between 0.3~keV and 1.5~keV that is divided into 24 energy bins with a bin width of 0.05~keV. The assumed spin frequency of the neutron star is 200~Hz and the period is divided into 32 phase bins. The total exposure time is assumed to be $10^6$s that is a typical value for NICER observations. We calculate the flux using our semi-analytic model~\citep{Zhao2024}, consider the interstellar extinction and the instrument response to get the final observed counts of photons, and add Gaussian noise to each energy and phase bin. For the instrument response, we adopt the effective area and response matrix of NICER and eXTP to generate two sets of synthetic data. For NICER, the total number of photons is $1.43\times 10^6$ that is the typical value of NICER observations. For eXTP, which has a better effective area and response matrix, this value becomes $4.41\times 10^6$, about three times the total counts of the synthetic NICER data. We fit our semi-analytic model to the synthetic pulse-profiles to estimate the uncertainty of the spot colatitude measurement. For each set of synthetic data, we perform Bayesian parameter inference using a Metropolis-Hastings Markov-Chain Monte Carlo (MCMC) algorithm with steps drawn from Gaussian distributions and no tempering. Our algorithm is loosely based on \texttt{MARCH} \citet{Psaltis2022}. We use flat-top distributions for the priors over the model parameters. For a given model $\cal M$ and parameters $\Theta$, we use the Gaussian likelihood function:
\begin{equation}
{\cal L}({\cal D}| {\cal M}, \Theta)=-\sum _i \frac{(D_i-M(\Theta)_i)^2}{\sigma_i^2},
\end{equation}
where $D_i$ and $M_i$ is the counts of photons in the ith energy and phase bin given by the data and our model, $\sigma_i$ is the standard deviation of the Gaussian error. The 68-th and 95-th percentiles of the marginalized posterior distributions for both synthetic NICER data and eXTP data are shown in Fig~\ref{fig:fig1}. 

The marginalized posterior over $\alpha$ has a width of about $0.2^\circ$ and $0.1^\circ$ for NICER and eXTP respectively. If we double the total exposure time, the uncertainty of $\alpha$ for eXTP data decreases to about $0.08^\circ$, but still in the same order of magnitude.

\section{Synthetic Data and Parameter Inference Results for Radio Pulse-Profiles}\label{sec:radio}
We synthesize the data of a radio telescope receiving the signal of a radio pulsar. The parameters of the radio telescope and the pulsar are listed in Table~\ref{table:paras}. RVM contains four parameters: the colatitude of the magnetic axis (the magnetic inclination angle) $\alpha_{\mathrm{M}}$, the colatitude of the line-of-sight (the observer inclination angle) $\zeta_{\mathrm{M}}$, the phase offset $\phi_{0,\mathrm{M}}$, and the PA offset $\psi_{0,\mathrm{M}}$. We would like to know that if the pulsar radio signal is polarized according to RVM, how well we could constrain $\alpha_{\mathrm{M}}$ by fitting typical pulsar data with RVM.
\begin{table*}[h!]
    \caption{Parameters}                
    \label{table:paras}   
    \centering                        
    \begin{tabular}{c l l}     
    \hline\hline               
    Quantity & Set value & Meaning\\         
    \hline                      
    Telescope parameters & \\
    \hline
    $G_{\mathrm{sys}}$ & 16 K/Jy & Gain\\
    $T_{\mathrm{sys}}$ & 25 K & System temperature \\
    $\Delta \nu$ & 400 MHz & Bandwidth \\
    \hline
    Pulsar parameters & \\
    \hline
    $P_{0}$ & 0.04 s & Spin period \\
    $W_{\mathrm{g}}$ & 0.002 s & Pulse width \\
    $S_{\nu,\mathrm{p}}$ & 20 mJy & Peak flux density \\
    $\alpha_{\mathrm{M}}$ & see text & Magnetic inclination angle \\
    $\zeta_{\mathrm{M}}$ & see text & Observer inclination angle \\
    $\phi_{0,\mathrm{M}}$ & 0.4 & Phase offset (phase of pulse peak) \\
    $\psi_{0,\mathrm{M}}$ & $\pi/6$ & PA offset \\
    \hline
    Observation parameters & \\
    \hline
    $\phi$ & from 0 to 1 & Pulse phase \\
    $\tau_{\mathrm{int}}$ & 3600 s & Integration time \\
    $N_{\mathrm{bin}}$ & 1024 & Number of bins \\
    \hline
    \end{tabular}
    \caption{In $\alpha_{\mathrm{M}}$, $\zeta_{\mathrm{M}}$, $\phi_{0,\mathrm{M}}$, and $\psi_{0,\mathrm{M}}$, $_{\mathrm{M}}$ means ``modeled''. The telescope parameters are taken from FAST performance paper~\citep{2020fast}.}
    \end{table*}

\par Assuming the radio pulse profile is described with a Gaussian function, we write the intensity (Stokes $I$) of profile (without noise) as
\begin{equation}
    I_{\mathrm{w}}(\phi) = G_{\mathrm{sys}} S_{\nu,\mathrm{p}} \exp{\left(-\dfrac{(\phi-\phi_{0,\mathrm{M}})^{2}P_{0}^{2}}{2W_{\mathrm{g}}^{2}}\right)},
    \label{eq:profile}
\end{equation}

\noindent $I_{\mathrm{w}}$ has the same dimension as temperature. For simplicity, we assume that the pulse profile is purely linearly polarized, and then $I^{2}=L^{2}=Q^{2}+U^{2}$. Therefore, Stokes $Q$, $U$, and $V$ could be expressed as
\begin{equation}
\left\{
\begin{aligned}
    Q_{\mathrm{w}}(\phi) &= I(\phi)\cos(2\psi_{M}(\phi)), \\ U_{\mathrm{w}}(\phi) &= I(\phi)\sin(2\psi_{M}(\phi)),\\ V_{\mathrm{w}}(\phi) &=0,
\end{aligned}
\right.
\label{eq:QU}
\end{equation}
\noindent where $\psi_{M}$ is the modeled PA at each phase. $\psi_{M}$ satisfies the equation of RVM:~\citep{RVM1969}
\begin{equation}
    \psi_{M}(\phi) = \tan^{-1} \dfrac{\sin\alpha_{\mathrm{M}}\sin(2\pi(\phi-\phi_{0,\mathrm{M}}))}{\sin\zeta_{\mathrm{M}}\cos\alpha_{\mathrm{M}}-\cos\zeta_{\mathrm{M}}\sin\alpha_{\mathrm{M}}\cos(2\pi(\phi-\phi_{0,\mathrm{M}}))}.
    \label{eq:RVM}
\end{equation}
The noise is introduced with the ideal radiometer equation~\citep[e.g.,][]{radiobook} 
\begin{equation}
    I_{n} = \dfrac{T_{\mathrm{sys}}}{\sqrt{\tau_{\mathrm{int}}\Delta \nu / N_{\mathrm{bin}}}}.
    \label{eq:radiometer}
\end{equation}
The noise added to each pulse phase is drawn from a Normal distribution with its mean value equal to zero and its standard deviation equal to $I_{n}$. We assume that the noise level of different Stokes parameters are the same, then the finalized Stokes parameters are
\begin{equation}
		\left\{
			\begin{aligned}
				I(\phi)&=I_{\mathrm{w}}(\phi)+\mathcal{N}(0,I_{n}),\\
				Q(\phi)&=Q_{\mathrm{w}}(\phi)+\mathcal{N}(0,I_{n}),\\
                U(\phi)&=U_{\mathrm{w}}(\phi)+\mathcal{N}(0,I_{n}),\\
                V(\phi)&=V_{\mathrm{w}}(\phi)+\mathcal{N}(0,I_{n}).\\
			\end{aligned}
			\right.
			\label{eq:bxby}
\end{equation}

We use $Q$ and $U$ for Bayesian inference of parameters $\alpha$, $\zeta$, $\phi_{0}$, and $\psi_{0}$. For a pair of arrays $\{Q(\phi)\}$ and $\{U(\phi)\}$ obtained from synthetic data, the marginalized likelihood function could be written as (refer to the Supplementary Materials of~\citealp{2019desvignes})

\begin{equation}
\begin{aligned}
    P(\mathcal{D}|\mathcal{M},\Theta) \propto & \exp\left(-\frac{1}{2}\sum_{\phi}\frac{[L'_{\phi}\cos(2\psi_{M}(\phi))-Q(\phi)]^{2}}{\sigma_{Q}^{2}(\phi)}\right. \\
    & \qquad \left.-\frac{1}{2}\sum_{\phi}\frac{[L'_{\phi}\sin(2\psi_{M}(\phi))-U(\phi)]^{2}}{\sigma_{U}^{2}(\phi)}\right).
\end{aligned}
\label{eq:likelihood}
\end{equation}

\noindent where $\sigma_{Q}(\phi)$ and $\sigma_{U}(\phi)$ are measurement errors of $Q(\phi)$ and $U(\phi)$ respectively. The errors are derived by calculating the standard deviation of $Q(\phi)$ and $U(\phi)$ in the baseline phase range (we choose phase bin 800 - 1024 as the baseline). $L'_{\phi}$ could be written as

\begin{equation}
    L'_{\phi} = \dfrac{\dfrac{Q(\phi)\cos(2\psi_{M}(\phi))}{\sigma_{Q}^{2}(\phi)}+\dfrac{U(\phi)\sin(2\psi_{M}(\phi))}{\sigma_{U}^{2}(\phi)}}{\dfrac{\cos^{2}(2\psi_{M}(\phi))}{\sigma_{Q}^{2}(\phi)}+\dfrac{\sin^{2}(2\psi_{M}(\phi))}{\sigma_{U}^{2}(\phi)}}.
    \label{eq:L}
\end{equation}

As for the prior probability distribution function (PDF) $P(\mathcal{M},\Theta)$, we set $\phi_{0}$ to be uniform within (0.3, 0.5), and $\psi_{0}$ to be uniform within (0, $\pi$). $\alpha$ and $\zeta$ are sampled to obey that $\sin\alpha$ and $\sin\zeta$ are uniform within (0, 1). We used the \texttt{python} package \texttt{emcee}~\citep{emcee} (version: 3.1.6) for Markov Chain Monte Carlo (MCMC) computations.

\par The data points that we use for Bayesian analysis satisfy $\sigma_{\psi} < 5^{\circ}$, to ensure high-enough signal-to-noise ratio of $Q$ and $U$. To scan the parameter space, we set $\alpha_{\mathrm{M}}$ equal to (10, 29, 50, 70, 90) degrees, $\zeta_{\mathrm{M}}$ equal to (11, 31, 51, 71, 91) degrees. We generate synthetic data with those 25 pairs of ($\alpha_{\mathrm{M}}$, $\zeta_{\mathrm{M}}$), and run Bayesian analysis to infer their ``observed''  ($\alpha_{\mathrm{M}}$, $\zeta_{\mathrm{M}}$) pairs. The $\phi_{0,\mathrm{M}}$ and $\psi_{0,\mathrm{M}}$ are fixed to their values in Table~\ref{table:paras}. We set 30000 steps for running and we omitted the first 20000 steps as burn-in steps, applicable to all cases.

\begin{figure}[h]
\centering
\begin{minipage}{0.9\linewidth}
\centering
\includegraphics[width=0.95\linewidth]{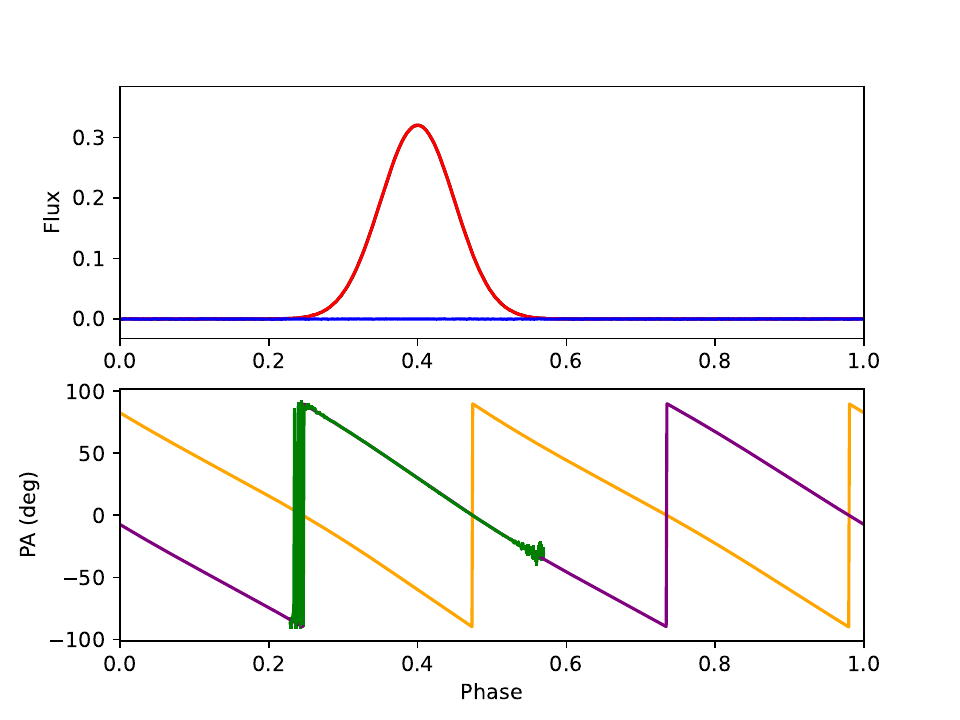}
\end{minipage}
\begin{minipage}{0.9\linewidth}
\centering
\includegraphics[width=0.98\linewidth]{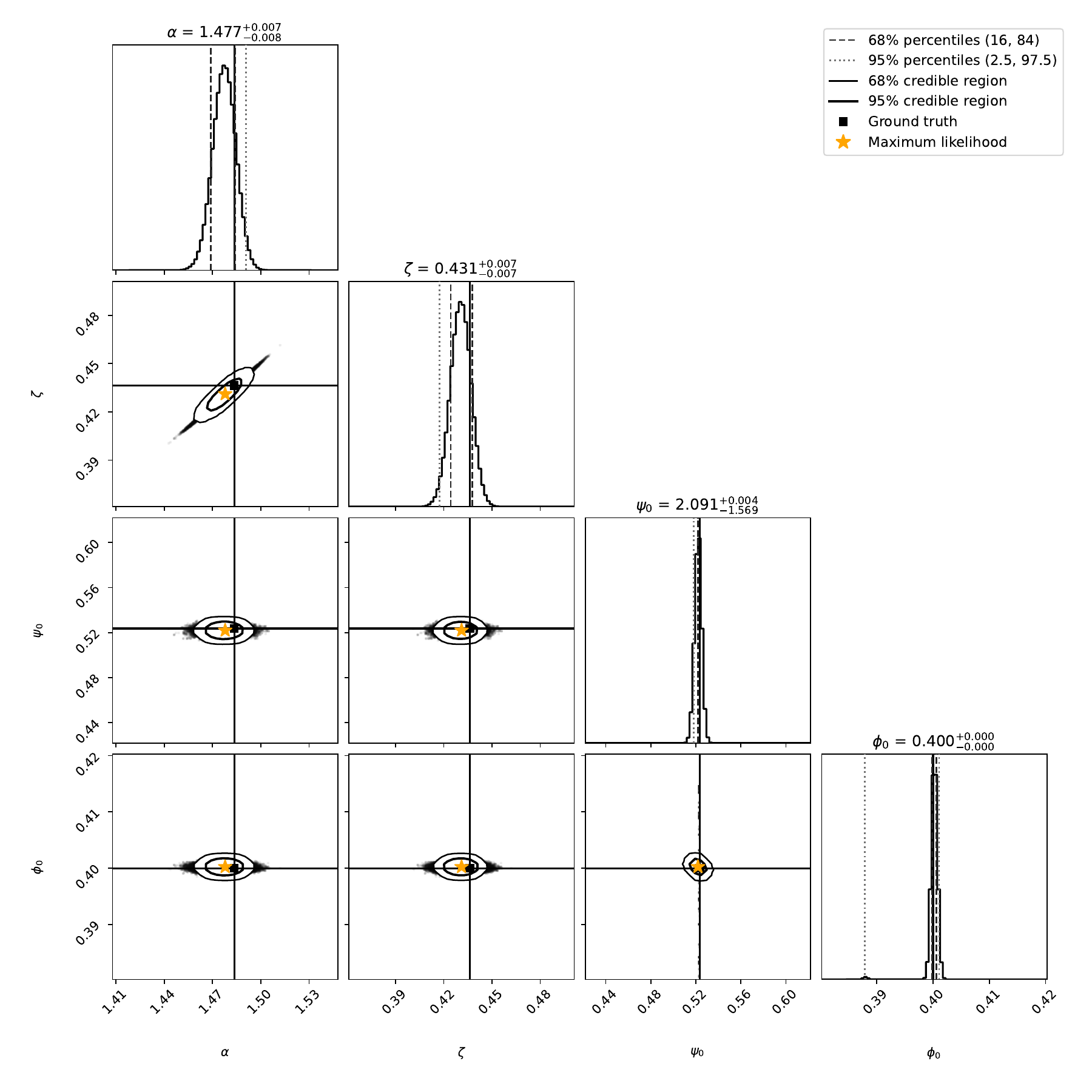}
\end{minipage}
    \caption{A synthetic radio pulse profile (left) and the posterior PDF of RVM fitting of this profile (right). Left: Black line shows the total intensity ($I$); red line shows the linear polarization intensity ($L=\sqrt{Q^{2}+U^{2}}$); blue line shows the circular polarization intensity ($V$). The horizontal axis is the pulse phase (0 to 1 in a period). Green dots in the PA panel with errorbars are the polarization position angles (PA, $\psi=0.5\arctan(U/Q)$). The orange curve in the PA panel is the RVM curve calculated with maximum-likelihood parameters inferred from Bayesian analysis. The purple curve in the PA panel is the orange curve shifted by 90 degrees. Right: the blue curves are marginal PDFs of parameters and the grey contours are 2D PDFs of each two parameters. The contour profiles mark the credibility regions of probabilities 68\% and 95\%. The orange stars represent the maximum-likelihood parameters. The black square with vertical and horizontal solid lines represent the reference parameters set for the synthetic data.}
    \label{fig:result_1}
\end{figure}

An example of the Bayesian analysis result is shown in Figure~\ref{fig:result_1}. The fitting is generally successful. Sometimes the RVM curve with maximum-likelihood parameters is shifted from the PA curve by 90 degrees, because the $L'_{\phi}$ defined in Equation~\ref{eq:L} could be positive or negative. A 90-degree shift of PA means a reverse of sign in $Q$ and $U$. 

To find out if the goodness of fitting relies on $\alpha_{\mathrm{M}}$ and $\beta_{\mathrm{M}}$, we compare the errors of inferred parameters of different pairs of ($\alpha_{\mathrm{M}}$, $\beta_{\mathrm{M}}$). The quoted parameter values correspond to the median of the posterior distributions, while the uncertainties represent the 16th and 84th percentiles (i.e., the central 68\% credible intervals) derived from the MCMC samples. The upper and lower uncertainties could be written as

\begin{equation}
    \Delta\alpha_{+} = \alpha_{84}-\alpha_{50}\quad \Delta\alpha_{-} = \alpha_{50}-\alpha_{16}.
\end{equation}

\begin{figure}[htbp]
\centering
\includegraphics[width=8.5cm]{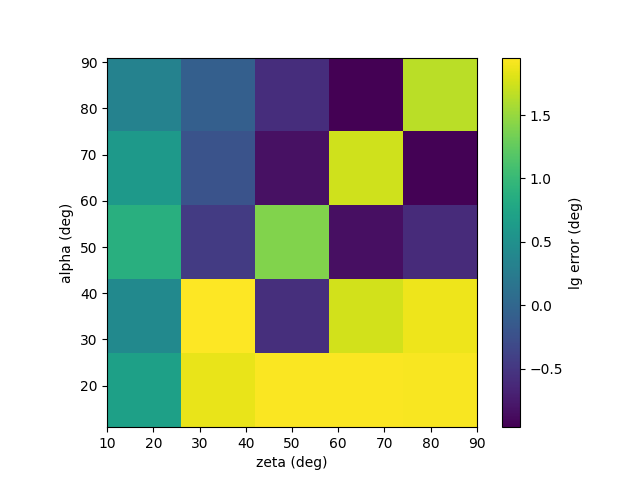}
\caption{The fitting errors of magnetic inclination angle $\alpha$ of different pairs of input $(\alpha_{\mathrm{M}}, \zeta_{\mathrm{M}})$. The colors represent the magnitude of errors in the unit of degree.}
\label{fig:sigma_arr}
\end{figure}

\noindent It is the same for $\zeta$. We take $\bar{\Delta\alpha} = (\Delta\alpha_{+}+\Delta\alpha_{-})/2$, and plot $\bar{\Delta\alpha}$ of ($\alpha_{\mathrm{M}}$, $\beta_{\mathrm{M}}$) pairs. The result is shown in Figure~\ref{fig:sigma_arr}.

\section{Discussion}
The RVM fit constrains the focus point of polarization orientations. It is usually though to be the center of the plasma flow. However, the polarization state of the radio emission changes as it propagates because the evolution of wave modes during wave propagation in the magnetosphere strongly depends on magnetic field orientations. Thus, we propose a novel method to identify whether the focus point of polarization orientations is consistent with the center of the plasma flow, by comparing the colatitude of the hot spot center, as inferred from X-ray pulse-profile modeling, with the magnetic inclination angle obtained from a RVM fit to radio data. The X-ray measurement traces the center of the in-falling particle flow onto the polar cap, while the RVM fit constrains the focus point of polarization orientations. We conclude that comparing the two independent angle measurements is a promising avenue for probing propagation effects in pulsar magnetospheres. While current NICER observations may lack the precision for this test, future missions like eXTP, with its enhanced effective area and spectral resolution, could provide the necessary accuracy to achieve this goal. Thus, this multi-wavelength approach may offer a powerful tool to understand the complex physics of pulsar emission and magnetospheric propagation.

Our fit results of synthetic thermal X-ray pulse-profiles show that in the most ideal situation with very low background and proper neutron star parameters, the minimal uncertainty of the spot colatitude measurement is about $0.1^\circ$ for the eXTP instrument response. The uncertainty of RVM measurement for pulsars with the same set parameters is also about $0.1^\circ$. This means the uncertainty of the deviation between these two angles is $\sigma_{total}\approx0.141^\circ$. For normal pulsars, the deviation between the plasma current center and the open magnetic field center is about $1^\circ$ in colatitude according to the 3D particle-in-cell simulation~\citep{Jan2026}. Thus this multi-messenger comparison maybe useful to distinguish the plasma current center from the polarization center. However, whether the measurement errors can be controlled to this extent in real observations requires further research. In addition, for millisecond pulsars that usually have larger polar caps, this comparison becomes more possible. Although measurement errors of the thermal X-ray pulse-profile and RVM measurements do not vary significantly with frequency, the size of the polar cap increases with increasing frequency. According to~\citet{Gralla2017}, for pulsars with large magnetic inclination angles, the center of current center is close to the boundary of the polar cap region. Thus, this deviation will be similar to the half opening angle of the polar cap, which is several degrees for millisecond pulsars. Although for most millisecond pulsars multi-polar magnetic fields are not negligible within the polar cap region. Observation shows that there is still a small part of millisecond pulsars suitable for RVM~\citep{Liu2025}.

Except for the hotspot modeling and radio RVM fitting mentioned in Section~\ref{sec:xray} and Section~\ref{sec:radio}, the X-ray polarimetry of normal pulsars and millisecond pulsars provides a third way of measuring pulsar radiation geometry. For thermal X-ray pulsars, the X-ray polarization is shaped by the vacuum birefringence effect above the neutron star surface~\citep{heyl2000}, and thus also follows RVM~\citep{meszaros1988}. This scenario has been tested by polarimetry of accretion-powered X-ray pulsars (refer to review paper e.g.~\citealp{ptf2024}). 

The RVM of X-ray pulsars results purely from propagation effect, and should depends only on the magnetic field configuration above the stellar surface. For high-precision X-ray polarimetry researches with future missions like eXTP, we may compare the radiative geometry measured by hot-spot modeling, radio RVM, and X-ray RVM. If the radio RVM result is the same as the X-ray RVM result but is different from the hot-spot modeling result, we may conclude that the conventional radio RVM is saved by the propagation effects.

\bibliographystyle{apj}
\bibliography{MS}
\end{document}